# Characterizing global tropical cyclone events of 2024

Lingke Jiang[1*], G. Brooke Anderson[2], Yanran Li,[1] Xiao Wu[1], Victoria D. Lynch[3], Robbie M. Parks[3]

[1]Department of Biostatistics, Columbia University Mailman School of Public Health, New York, NY, USA.

[2]Department of Environmental & Radiological Health Sciences, Colorado State University, Fort Collins, CO, USA.

[3]Department of Environmental Health Sciences, Columbia University Mailman School of Public Health, New York, NY, USA.

* Correspondence to: lj2575@cumc.columbia.edu

**Full impact assessment of tropical cyclones each year requires a comprehensive sociodemographic analysis. We evaluated sociodemographic characteristics of tropical cyclone-impacted regions during the 2024 calendar year in recent historical context of 1980-2024. In 2024, tropical cyclone-force wind affected an estimated 429,902,820 people (5.5% of global population), and hurricane-force wind an estimated 59,672,600 people (0.8%). Our findings provide a global context for tropical cyclones to better guide resilience and recovery efforts.**

Tropical cyclones, the strongest of which are also known as hurricanes, typhoons or cyclones, have devastating and far-reaching effects on society.[1–5] Timely quantifications of who, where, and when recent tropical cyclones have impacted regions worldwide in the past 40 years is critical for infrastructure recovery and mitigating impacts on communities:[2,6] historical exposure data may facilitate resource allocation to areas with risk of exposure; indeed, as shown by a recent comparative analysis of Hurricane Milton and Helene, areas with prior experience in tropical cyclone-related evacuations exhibited greater adaptive capacity.[7]

Here, we evaluated the exposure and sociodemographic characteristics of regions most impacted by tropical cyclone exposures during the 2024 calendar year, and place in the context of past 40 years' tropical cyclone activities. For all storms that occurred during 2024 globally, we used tracking data from International Best Track Archive for Climate Stewardship (IBTrACS) to model local windfields at second-level administrative unit (ADM2, US county-equivalent; referred to as "area" hereafter) level. We defined exposure to tropical cyclone- ($\geq$ 17.5 m/s) and hurricane-force ($\geq$ 32.9 m/s) when the peak sustained wind that day in the population center of the area associated with the tropical cyclone reached or exceeded these thresholds at the point of closest approach.[8,9] We combined modeled wind fields with gridded population estimates from Global Human Settlement Layer (GHSL) to generate estimated exposed population.[10,11] We then compared 2024 person-day exposure to prior 44 years of tropical cyclone records (1980-2024). Further, we assessed the relationship between person-days and the Global Gridded Relative Deprivation Index (GRDI), a composite measure of deprivation based on child dependency ratio, infant mortality rate, subnational human development index, as well as the recent average and the historical trend of nighttime lights intensity.[12,13] Finally, we ranked the most affected areas by tropical cyclone- and hurricane-force person-day exposure and identified storms that contributed the most to global exposure. All analyses and visualizations were stratified by World Health Organization (WHO) regional offices—African Region, Region of the Americas, Eastern Mediterranean Region, European Region, South-East Asian Region, and Western Pacific Region—as defined by *World Health Statistics 2020*.[14] A map illustrating the regional divisions is included in Supplementary Figure 1.

Globally, during 2024, there were 94 uniquely identified tropical storms recorded by IBTrACS, of which 39 contributed to 507,926,767 tropical cyclone-force person-days in 3,091 areas; 16 contributed to 60,228,900 hurricane-force person-days in 712 areas (Figure 1). In total, 429,902,820 people (5.5% of global population) were estimated to have been exposed to tropical cyclone-force winds; 59,672,600 people (0.8% of global population) were estimated to have been exposed to hurricane-force winds. Tropical cyclone-exposed areas globally were disproportionately more deprived areas (56.5% with moderately high or high deprivation);

hurricane-exposed areas globally were disproportionately less deprived areas (54.1% with moderately low or low deprivation).

For tropical cyclone-force WHO region-level exposure, the Western Pacific (298,701,690 person-days) ranked highest, followed by South-East Asia (118,107,940 person-days), the Americas (75,697,540 person-days), Europe (8,444,540 person-days), and Africa (6,973,980 person-days) (Figure 1A). Tropical cyclone-exposed areas were disproportionately less deprived areas in the Western Pacific (67.0% with moderately low or low deprivation); disproportionately more deprived areas in South-East Asia (88.9% with moderately high or high deprivation); disproportionately more deprived areas in the Americas (72.8% with moderately high or high deprivation); disproportionately less deprived areas in Europe (56.3% with moderately low or low deprivation); and disproportionately more deprived areas in Africa (95.1% with moderately high or high deprivation) (Figure 1B).

For hurricane-force WHO region-level exposure, the Western Pacific (49,645,170 person-days) ranked first, followed by the Americas (9,891,650 person-days), and Africa (692,080 person-days) (Figure 1C). Hurricane-exposed areas were disproportionately less deprived areas in the Western Pacific (73.6% with moderately low or low deprivation); were disproportionately more deprived areas in the Americas (70.9% with moderately high or high deprivation); and were disproportionately more deprived areas in Africa (80% with moderately high or high deprivation) (Figure 1D). No area in South-East Asia and Europe was exposed to hurricanes in 2024, despite substantial tropical cyclone-force exposure.

Placing 2024 in a historical context, for tropical cyclone-force exposure, global exposure (429,902,820 person-days) ranked 12[th] since 1980 (Figure 2A), while regionally Western Pacific (298,701,690 person-days) was ranked 24[th], Americas (75,697,540 person-days) was ranked 7[th], South-East Asia (118,107,940 person-days) was ranked 5[th], Europe (8,444,540 person-days) was ranked 4[th], and Africa (6,973,980 person-days) was ranked 13[th].

The top ten areas most affected by tropical cyclone-force exposure in 2024 were all located in the Western Pacific or South-East Asia (Figure 2B). Major urban centers in China (Shanghai; 17,495,090 person-days), Bangladesh (e.g., Dhaka; 16,282,930 person-days), India (e.g., South 24 Parganas; 12,357,410 person-days), and the Philippines (e.g., Isabela; 1,912,970 person-days) were areas with the highest person-day tropical cyclones-force exposure in 2024. All top ten most affected areas (apart from mega-cities Shanghai, Dhaka, and North 24 Parganas), had relatively moderate to high levels of deprivation (e.g., Isabela in Philippines (high deprivation)).

Cyclone Remal caused the most tropical cyclone-force person-day exposure (101,204,150 person-days) in 2024 (Figure 2C), primarily affecting areas in Bangladesh (e.g., Dhaka, 16,282,930 person-days) and India (e.g., South 24 Parganas, 12,357,410 person-days), compared with Typhoon Yagi in second (50,863,469 person-days). Except for Hurricane Helene (ranked 8[th]), which affected Mexico and the USA, all other storms on the top 10 list impacted regions of Western Pacific and South-East Asia, with repeat exposure in China (90,242,460 person-days including from Typhoon Yagi, Typhoon Gaemi, Typhoon Kong-Rey, Typhoon Bebinca), the Philippines (85,027,340 person-days including from Typhoon Yagi, Typhoon Kong-Rey, Typhoon Ewiniar, Typhoon Man-Yi), Japan (36,285,210 person-days including from Typhoon

Gaemi, Typhoon Bebinca, Typhoon Ampil), India (59,178,750 person-days including from Cyclone Remal, Cyclone Dana), and Taiwan (51,854,580 person-days including from Typhoon Gaemi, Typhoon Kong-Rey).

Placing 2024 in a historical context, for hurricane-force exposure, global exposure (60,228,900 person-days) ranked 10th since 1980 (Figure 2D), while regionally Western Pacific (49,645,170 person-days) was ranked 13th, Americas (9,891,650 person-days) was ranked 5th, and Africa was ranked 14th (692,080 person-days).

The top ten areas most affected by hurricanes in 2024 were in the Western Pacific or the Americas (Figure 2E). Urban areas in the Philippines (e.g., Isabela, 1,912,970 person-days), China (Haikou, 1,482,430 person-days), and USA (Hillsborough County, 1,454,910 person-days) were areas with the highest hurricane-force person-day exposure. Among the top ten areas with highest person-day exposure, the four in the Philippines had moderately high or high levels of deprivation while the other six had moderately low or low levels of deprivation (e.g., Haikou in China).

Typhoon Yagi caused the most hurricane-force person-days exposure (22,317,720 person-days) in 2024 (Figure 2F), primarily affecting areas in China (e.g., Haikou, 1,482,430 person-days) and Vietnam (e.g., Hoang Mai, 477,970 person-days), compared to Typhoon Gaemi in second (12,207,420 person days). In the Western Pacific, the Philippines (7,332,560 person-days including from Typhoon Man-Yi, Typhoon Toraji, Typhoon Usagi), Taiwan (18,582,660 person-days, including from Typhoon Gaemi, Typhoon Kong-Rey), Japan (1,412,230 person-days, including from Typhoon Gaemi, Typhoon Shanshan) were repeatedly affected. In the Americas, the USA (5,347,460 person-days including from Hurricane Milton, Hurricane Usagi) and Mexico (1,133,290 person-days including from Hurricane Milton, Hurricane Usagi) were repeatedly affected.

Global-level tropical cyclone-force and hurricane-force person-day exposure has increased over time. Globally, each additional year is associated with an increase of 6,926,350 tropical cyclone-force person-days ($p < 0.001$) and 873,530 hurricane-force person-days ($p < 0.001$). However, when normalizing by population size, per-capita tropical cyclone- or hurricane-force person-day exposure shows no clear trend over time ($p > 0.05$).

**Methods**

For all storms that occurred during 2024, we obtained tracking data from IBTrACS to model local windfields at ADM2 level globally,[8,9] with full time and space coverage of the study period. We generated peak sustained wind at each ADM2 population centroid for any storm passing within a 250 km radius via the Willoughby wind speed model, detailed elsewhere.[15] As in previous work,[2,3] we defined tropical cyclone-force exposure as days when the peak sustained wind in the population center of the ADM2 associated with the tropical cyclone at the point of closest approach reached or exceeded 34 knots (63 km / hour, 17.1 m/s; gale force wind on Beaufort scale), and hurricane-force exposures as all days with sustained winds greater than or equal to 64 knots (119 km / hour, 33.1 m/s). We then obtained 1km-resolution population grids from Global Human Settlement Layer (GHSL)'s quinquennial releases (1980-2020),[10,11] aggregated them to ADM2 units and linearly interpolated to produce annual values; we assumed the 2020 population data for years that succeed 2020. Further, we combined the daily exposure estimates with population data to generate person-days of exposure, defined as the population of an ADM2 at year of exposure multiplied by the number of days they were exposed. Additionally, we obtained the Global Gridded Relative Deprivation Index (GRDI, v1; 2010-2020) from NASA Socioeconomic Data and Applications Center (SEDAC),[12,13] aggregated them to ADM2 level and merged with exposure data. GRDI values were classified into quartiles: 0-0.25 (low deprivation), 0.2501-0.5000 (moderately low), 0.5001-0.7500 (moderately high), and 0.7501-1.0 (high). To assess temporal trends, we performed linear regression with year as the independent variable and global tropical cyclone- and hurricane-force person-day exposure as the dependent variables; per-capita trends were analyzed by dividing annual person-day exposure by global population before regression, with statistical significance set at p < 0.05. For each analysis and visualization, we stratified the results by World Health Organization (WHO) region to examine geographic variations in exposure and deprivation.

**Data availability**

Data used for this analysis are available via https://github.com/sparklabnyc/global_tc_2024/tree/01_data. The data used in this study were created from the following datasets. Tropical cyclone best track data during 1980-2024 are freely available at https://www.ncei.noaa.gov/products/international-best-track-archive; global gridded population data at 1km resolution during 1980-2020 are available quinquennially at https://human-settlement.emergency.copernicus.eu/download.php?ds=pop; global gridded deprivation index data at 1km resolution in 2020 are available at https://www.earthdata.nasa.gov/data/catalog/sedac-ciesin-sedac-pmp-grdi-2010-2020-1.00. An interactive shiny app is available via https://lincolej.shinyapps.io/G-TROPIC_2024/.

**Code availability**

All code to reproduce this work are freely available at https://github.com/sparklabnyc/global_tc_2024/.

**Author contributions**

G.B.A. and R.M.P designed research; G.B.A., L.J., and R.M.P. performed research; G.B.A. and V.D.L contributed analytic tools; G.B.A., L.J., R.M.P., and V.D.L. analyzed data; and L.J. and R.M.P wrote the paper with assistance from G.B.A., V.D.L., Y.L. and X.W.

**References**

1. Peduzzi, P. *et al.* Global trends in tropical cyclone risk. *Nat. Clim. Change* **2**, 289–294 (2012).

2. Parks, R. M. *et al.* Association of Tropical Cyclones With County-Level Mortality in the US. *JAMA* **327**, 946–955 (2022).

3. Parks, R. M. *et al.* Tropical cyclone exposure is associated with increased hospitalization rates in older adults. *Nat. Commun.* **12**, 1545 (2021).

4. Parks, R. M. *et al.* Short-term excess mortality following tropical cyclones in the United States. *Sci. Adv.* **9**, eadg6633 (2023).

5. Huang, W. *et al.* Global short-term mortality risk and burden associated with tropical cyclones from 1980 to 2019: a multi-country time-series study. *Lancet Planet. Health* **7**, e694–e705 (2023).

6. Parks, R. M. & Guinto, R. R. Invited Perspective: Uncovering the Hidden Burden of Tropical Cyclones on Public Health Locally and Worldwide. *Environ. Health Perspect.* **130**, 111306 (2022).

7. Qing Yao *et al.* Adaptive mobility responses during Hurricanes Helene and Milton in 2024.

8. Kenneth, R., Howard, J., James, P., Michael, C. & Carl, J. International Best Track Archive for Climate Stewardship (IBTrACS) Project, Version 4. (2019) doi:10.25921/82ty-9e16.

9. Knapp, K. R., Kruk, M. C., Levinson, D. H., Diamond, H. J. & Neumann, C. J. The International Best Track Archive for Climate Stewardship (IBTrACS). (2010) doi:10.1175/2009BAMS2755.1.

10. Carioli, A., Schiavina, M., Freire, S. & MacManus, K. GHS-POP R2023A - GHS population grid multitemporal (1975-2030). (2023) doi:10.2905/2FF68A52-5B5B-4A22-8F40-C41DA8332CFE.

11. Pesaresi, M. *et al.* Advances on the Global Human Settlement Layer by joint assessment of Earth Observation and population survey data. *Int. J. Digit. Earth* **17**, 2390454 (2024).

12. Center for International Earth Science Information Network (CIESIN), C. U. Documentation for the Global Gridded Relative Deprivation Index (GRDI). https://doi.org/10.7927/xwf1-k532 (2022).

13. Earth Science Data Systems, N. Global Gridded Relative Deprivation Index (GRDI), Version 1 | NASA Earthdata. Earth Science Data Systems, NASA (2025).

14. *World Health Statistics 2020*. (World Health Organization, Geneva, 2021).

15. Anderson, G. B. *et al.* stormwindmodel: Model Tropical Cyclone Wind Speeds. (2018).

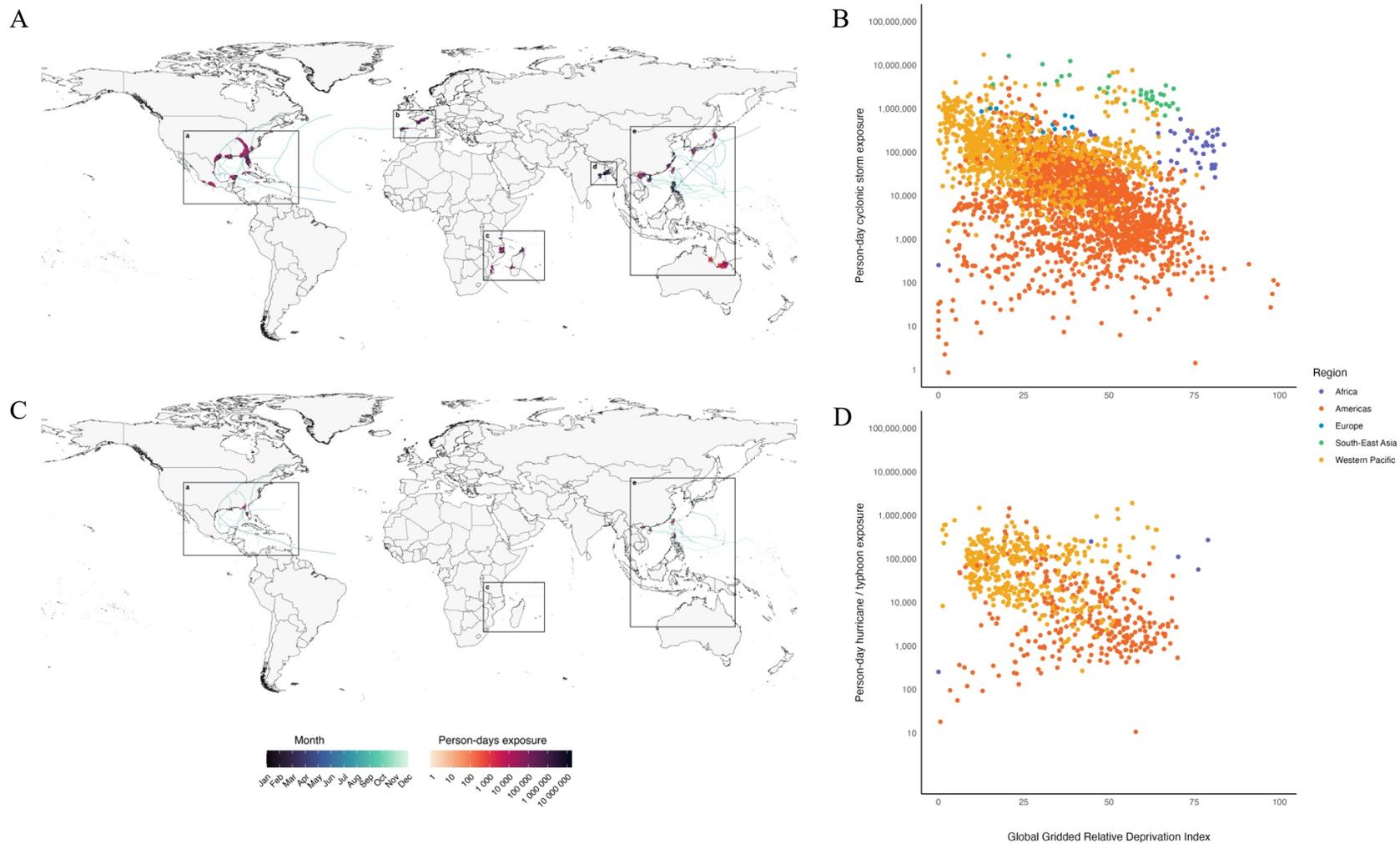

**Figure 1**. Global distribution of tropical cyclone and hurricane exposure. (A) Total person-day tropical cyclone exposure by ADM2. (B) Person-day tropical cyclone exposure against GRDI for each ADM2, stratified by WHO region. (C) Total person-day hurricane exposure by ADM2. (D) Person-day hurricane exposure against GRDI for each ADM2, stratified by WHO region.

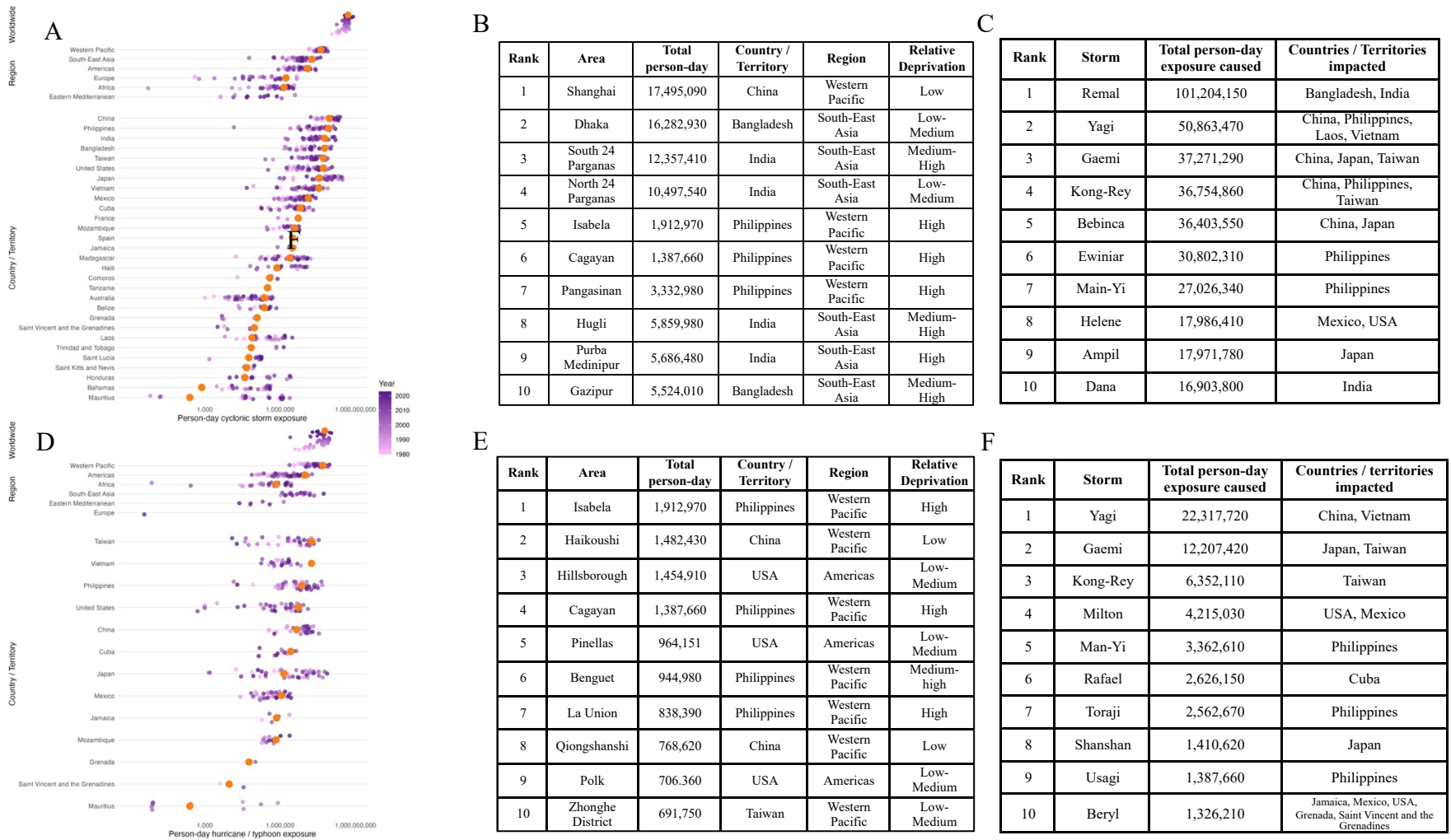

**Figure 2**. Global and regional patterns of tropical cyclone and hurricane exposure. (A) Global tropical cyclone exposure in 2024 in historical context (1980-2024), stratified by global total, WHO Region, and country. (B) Top 10 ADM2s with the highest tropical cyclone exposure in 2024. (C) Top 10 storms contributing the most to tropical cyclone exposure. (D) Global hurricane exposure in 2024 in historical context (1980-2024), stratified by global, WHO Region, and country. (E) Top 10 ADM2s with the highest person-day hurricane exposure. (F) Top 10 storms that contributing the most to hurricane exposure.

Supplementary Information

**Characterizing global tropical cyclone events of 2024**
Lingke Jiang, G. Brooke Anderson, Yanran Li, Xiao Wu, Victoria D. Lynch, Robbie M. Parks

**Supplementary Figure 1. World regions according to the World Health Organization.**

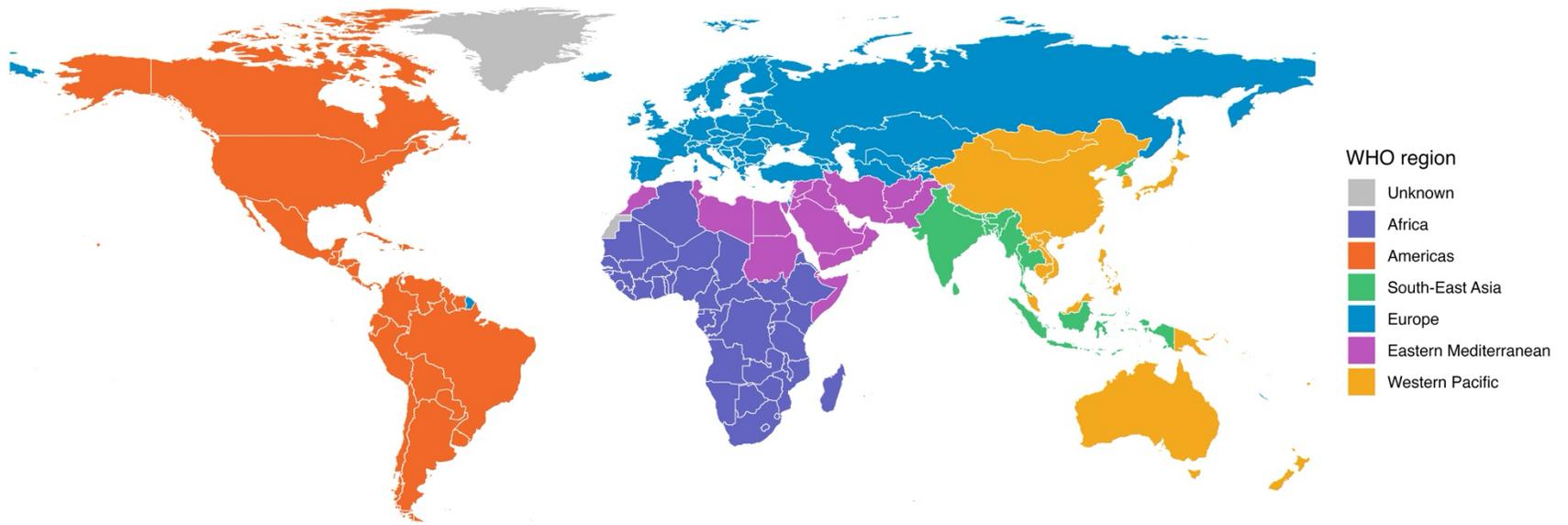

**Supplementary Figure 1: World regions according to the World Health Organization.**